\documentclass[aps,pra,amsfonts, amssymb, amsmath,groupedaddress,showkeys,reprint,superscriptaddress]{revtex4-2}
\usepackage{orcidlink}
\usepackage[american]{babel} % American English hyphenation
\usepackage{mathtools}
\usepackage{physics}
\usepackage{xfrac}
\usepackage{txfonts}
\usepackage{stmaryrd}
\usepackage[T1]{fontenc}
\usepackage{graphicx}
\usepackage{bm}
\usepackage{hyperref}

\usepackage{comment}
\newcommand{\corr}[1]{{\color{black}#1}}

\begin{document}
%\title{Monogamy of entanglement inspired protocol to quantify bipartite entanglement using spin squeezing}
\title{\corr{Monogamy-of-Entanglement-Inspired Protocol for Quantifying Bipartite Entanglement Using Spin Squeezing}}

\author{Diego Fallas Padilla}
\email{difa1788@colorado.edu}
\affiliation{JILA and Department of Physics, University of Colorado, 440 UCB, Boulder, CO, 80309, USA}
\affiliation{Center for Theory of Quantum Matter, University of Colorado, Boulder, CO 80309, USA}
\affiliation{Department of Physics and Astronomy, and Smalley-Curl Institute, Rice University, Houston, TX 77251-1892, USA}
\author{Mingjian Zhu}
\affiliation{Department of Physics and Astronomy, and Smalley-Curl Institute, Rice University, Houston, TX 77251-1892, USA}
\author{Han Pu}
\email{hpu@rice.edu}
\affiliation{Department of Physics and Astronomy, and Smalley-Curl Institute, Rice University, Houston, TX 77251-1892, USA}

\begin{abstract}
Quantum entanglement is an essential resource for quantum science and technology. However, entanglement detection and quantification, via typical entanglement measures such as linear entanglement entropy or negativity, can be a very challenging task. Here we propose a protocol to detect bipartite entanglement in a system of $N$ qubits inspired by the concept of monogamy of entanglement, \corr{where}, given a total system \corr{in a pure state} with some bipartite entanglement between two subsystems, subsequent unitary evolution and measurement of one of the subsystems may be used to quantify the entanglement between the two. To address the difficulty of detection, we propose to use spin squeezing to quantify the entanglement within the individual subsystem. Knowing that the relation between spin squeezing and some entanglement measures is not one-to-one, we give some suggestions on how a judicious choice of squeezing Hamiltonian can lead to better results in our protocol. For systems with a small number of qubits, we derive analytical results and show how our protocol can work optimally for GHZ states. For larger systems, we show how the accuracy of the protocol can be improved by a proper choice of the squeezing Hamiltonian. Our protocol presents an alternative for entanglement detection in platforms where state tomography is inaccessible or hard to perform. Additionally, the ideas presented here can be extended beyond spin-only systems to expand their applicability. 
\end{abstract}

\date{\today }
\maketitle

\section{Introduction}

Entanglement is a fundamental feature of quantum mechanics that has puzzled scientists since the early days of its formulation \cite{einstein1935can,schrodinger1935gegenwartige,bell1964einstein}. Apart from generating interesting philosophical discussions, quantum entanglement is an essential asset for applications in multiple quantum science fields such as quantum teleportation \cite{bennet1998,boschi,bouwmeester1997experimental}, quantum cryptography \cite{Pirandola:20}, quantum metrology \cite{giovannetti2011advances}, and quantum computation \cite{raussendorf2001one}.

%The tensor product space of two sub-Hilbert spaces possesses significantly more degrees of freedom compared to the simple products of states from the individual sub-spaces.

In general, we can think of the study of entanglement as divided into three questions: (i) How is entanglement generated? (ii) How is entanglement quantified? (iii) How is entanglement detected? In this work, we focus mostly on the second and third points but it is important to mention that, in the last couple of decades, great progress has been achieved in the generation of entanglement using photons \cite{hongoumandel,lu2007experimental,Pan2012}, trapped ions \cite{haffner2005scalable,leibfried2005creation,bohnet2016quantum,lu2019global}, and large ensembles of atoms in cavity QED setups \cite{SS2010,Haas2014,mcconnell2015entanglement}, among other platforms.

Regarding entanglement quantification, various entanglement measures and inequalities have been proposed in the past decades (for a complete review of them, see \cite{horodecki2009quantum}). For pure states, inequalities based on quantities such as entanglement entropy and linear entropy can be used to quantify the degree of entanglement in a system \cite{wei2003maximal}. In the case of mixed states, entanglement quantification is more complex due to the presence of classical correlations. A well-studied alternative is the use of quantum negativity, which is numerically accessible for both pure and mixed states  \cite{horodecki2001separability,zukowski1998quest}. Other measures of entanglement, such as the entanglement of formation \cite{bennett1996mixed} and relative entanglement entropy \cite{vedral1997statistical} stemming from concepts of distillable entanglement and entanglement cost, respectively, can be used to quantify entanglement in mixed states. %These measures can be hard to evaluate, even numerically, when the dimension of the Hilbert space is large. 

% From a quantum information perspective, entanglement is often characterized by entropic inequalities \cite{wehrl1978general}. % 
%quantify the amount of entanglement that can be extracted from a given state or the entanglement required to create a given state,% 

Concerning experimental entanglement detection, characterizing entanglement using the above measures can be difficult. Even in the case of pure states, obtaining the entanglement entropy would require quantum state tomography which is typically not easy to perform \cite{Rippe2008,takeda2021quantum} especially as the system size increases. An entanglement witness is defined as a measurable observable that can help distinguish between an entangled state and a separable one \cite{guhne2009entanglement}. For example, in a spin-only system, spin-squeezing parameters can be used to detect entanglement \cite{MA201189}. Spin-squeezing might not always quantify entanglement accurately \cite{Korbicz2005} \cite{wang2002pairwise}, \corr{but it is in general much easier to measure than the entanglement entropy. } Furthermore, spin-squeezed states are interesting in their own right, as they have extensive applications in quantum metrology \cite{Wineland1992,Pezze2018}. 

In this work, we explore the implementation of using spin-squeezing parameters of a subsystem to quantify the entanglement entropy of the full system, indirectly. This is particularly useful in situations where full-state tomography is impossible and manipulation and measurement of only one of the subsystems are available. As we show here, the relationship between spin squeezing in the subsystem and the entanglement of the full system is closely related to the monogamy of entanglement, a concept proposed and developed in the past decades \cite{PhysRevA.61.052306,wootters1998quantum,Calamet2017,Calamet2018}. 

The prototypical example of monogamy of entanglement involves three qubits A, B, and C. The monogamy property states that the entanglement between A and B will limit the entanglement between A and C, illustrating that the amount of information that A can share with other qubits is constrained. In particular, if A and B are maximally entangled, neither A nor B can share any entanglement with C. The statement can be extended to the case of $N$ qubits where A can only share a fixed amount of information with the other $N-1$ qubits \cite{osborne2006general}. In a recent paper \cite{ge2023faithful}, this property was generalized to the case where A, B, and C each describe a subsystem of an arbitrary tripartite quantum system in a pure state, for which the monogamy inequality is then interpreted geometrically in the form of a triangular relation. Other generalized versions of the monogamy criteria can be found in Ref.~\cite{zong2022}. 

\begin{figure}[t!]
\begin{centering}
\includegraphics[width=0.48\textwidth]{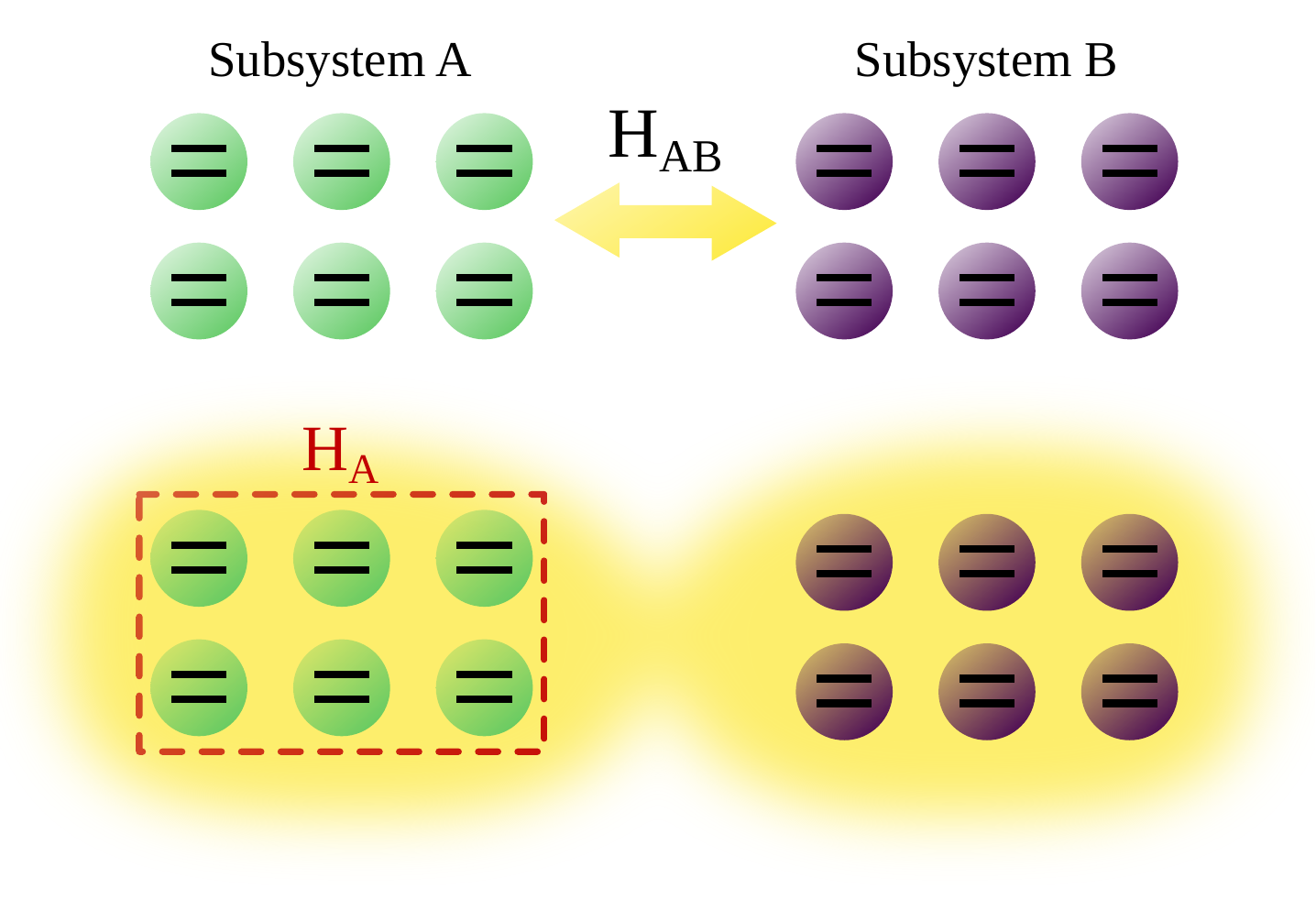}
\par\end{centering}
\caption{Schematics of the problem. Subsystem A contains $N_A$ two-level systems and subsystem B contains $N_B$ of them. First, through $H_{AB}$ the two subsystems get entangled. Then, subsequent evolution using $H_A$ entangles the elements within subsystem A without changing the bipartite entanglement between A and B.}
\label{Fig1}
\end{figure}

In this study, we are interested in \corr{utilizing} a generalization of the monogamy of entanglement concept, \corr{introduced initially in Ref.~\cite{Calamet2017,Calamet2018}}, which we schematically depict in Fig.~\ref{Fig1}. 
We consider a total system divided into two subsystems A and B, composed of $N_A$ and $N_B$ qubits, respectively. First, the entire system is subject to the action of Hamiltonian $H_{AB}$ which might create some bipartite entanglement between A and B. Subsequently, a Hamiltonian $H_A$ acts locally on subsystem A. The action of this Hamiltonian might create entanglement within the elements of A but its locality prevents any further modification of the bipartite entanglement between A and B. \corr{As we show later, the} entanglement generation within A \corr{is conditioned} by the pre-existing bipartite entanglement between A and B. \corr{Here, we address if} we can use the information obtained through the action of $H_A$ to learn about the bipartite entanglement between A and B. Moreover, we explore if these relationships can be studied in terms of easily measurable observables such as spin-squeezing parameters. 

This paper is organized as follows. First, we introduce different types of bipartite entanglement measures to be adopted in the paper in Section \hyperref[SecII]{II}. Then in Section \hyperref[SecIII]{III}, we \corr{introduce} the generalized monogamy property by investigating two simple cases where analytical expressions for the entanglement measures can be derived \cite{Calamet2017, Calamet2018}. Next, we introduce our general protocol for quantifying the entanglement between A and B through local operations and measurements on A in Section \hyperref[SecIV]{IV}. We then explore how spin-squeezing parameter measurements are related to bipartite entanglement in Section \hyperref[SecV]{V}. In Section \hyperref[SecVI]{VI},  we test how our protocol performs in systems with different numbers of spins for different typical Hamiltonians such as one-axis twisting (OAT) or two-axis twisting (TAT), and reflect on how to decide which of them might be more effective in different steps of our protocol. Finally, we conclude and discuss the limitations and advantages of the protocol in Section \hyperref[SecVII]{VII}.

\section{Bipartite Entanglement Measures} \label{SecII}
In this section, we introduce the bipartite entanglement measures we are going to use in this work. For all of them, we consider a bipartite system described by a total density matrix $\rho_{AB}$ and corresponding reduced density matrices $\rho_A$ and $\rho_B$. First, we introduce the concurrence, which is used for quantifying entanglement in generic states (pure or mixed).
For an arbitrary two-qubit state, the concurrence is equivalent to the entanglement of formation, which quantifies the amount of entanglement required to create such state \cite{bennett1996mixed,coffman2000distributed}, and it is defined mathematically as:
\begin{equation}
    C(\rho) = \text{max}(0, \mu_1 - \mu_2 - \mu_3 - \mu_4)\,.
    \label{cur}
\end{equation}
Where $\rho$ is the two-qubit density matrix. $\mu_j$ are the eigenvalues of the matrix $\Tilde{\rho} = \rho S \rho^* S$, with $S = \sigma_y \otimes \sigma_y$, and $\sigma_y$ being the conventional Pauli matrix acting on each qubit subspace. The eigenvalues are defined in descending order, namely, $\mu_1 \geq \mu_2 \geq \mu_3 \geq \mu_4$. If $C(\rho)=0$ the state is separable and if $C(\rho)=1$ the state is maximally entangled. 

Another measure we use for general states is the entanglement negativity. 
The positive partial transpose (PPT) criterion, or Peres-Horodecki criterion, is a necessary condition for separability which states that, for any separable state, the partial transpose of the composite density matrix, $\rho^{T_A}_{AB}$ (or equivalently  $\rho^{T_B}_{AB}$), is positive \cite{Peres1996,horodecki2001separability}. For systems of size $2\otimes2$ and $2\otimes 3$, the PPT condition is also proved to be sufficient. The entanglement negativity is a computable entanglement measure introduced to quantify the violation of the PPT condition and it is defined by:
\begin{equation}
    N_0(\rho) = \sum_{\lambda_i^{T_A}<0} |\lambda_i^{T_A}|\,,
    \label{neg}
\end{equation}
where $\lambda^{T_A}_i$ are the eigenvalues $\rho^{T_A}_{AB}$. For separable states $N_0(\rho)=0$ while $ N_0(\rho)>0$ indicates the presence of entanglement. The number of states with $N_0(\rho)>0$ can be used as an upper bound for the volume of entangled states in $N$-dimensional quantum systems \cite{zukowski1998quest}. For convenience of comparison, we introduce a normalized negativity $N(\rho)= N_0(\rho)/\max{(N_0(\rho))}$ such that the maximum available negativity for a system equals one.

Finally, we introduce the Tsallis entropy defined as: 
\begin{equation}
    T_q(\rho) = \frac{1-\text{Tr}(\rho^q)}{q-1}\,, \label{T_ent}
\end{equation}
with $q$ being a positive integer and $\rho$ being the density matrix of either the total system or a given subsystem. It can be shown that the inequality $T_{q}(\rho_{AB})-T_{q}(\rho_{A})\geq0$ holds for all separable states \cite{abe2001nonadditive}. If the composite system $\rho_{AB}$ is in a pure state such that the first term in the inequality vanishes, the bipartite entanglement can then be characterized by $T_q(\rho_A)=T_q(\rho_B)$. For $q=1,2$, $T_q$ reduces to Von-Neumann entropy and linear entropy, respectively. To compare the entropy of systems with different sizes, we adopt a normalized form of linear entropy:
\begin{equation}
    S_{L}(\rho) =\frac{N}{N-1}(1-\text{Tr}(\rho^2)), \label{l_ent}
\end{equation}
where $N$ is the dimension of the Hilbert space of $\rho$. In the following sections, we use the linear entropy to characterize the entanglement in pure states, and consequently we consider $\rho = \rho_A,\, \rho_B$ in Eq.~\eqref{l_ent}.

\section{The Generalized Monogamy of Entanglement} \label{SecIII}

In this section, we \corr{introduce} the generalized monogamy property for two specific cases. \corr{This concept was introduced theoretically in Ref.~\cite{Calamet2017,Calamet2018}, and later explored experimentally in Ref.~\cite{ExpMonogamy2020,ExpMonogamy2021}, where the entanglement between A and B is referred to as external entanglement, while the entanglement between different components of A is named internal entanglement. Here, we reintroduce some of the concepts derived in these prior works.}  
\corr{With a bipartition of system A into two subsystems $A_1,A_2$, the relation between the external and internal entanglement can be described by a monogamy inequality~\cite{Calamet2017}:
\begin{equation}
    {E}_{A_1,A_2} + E_{A,B}\leq E^{\max}_{A_1,A_2},\label{E_ineq} 
\end{equation}
where $E_{A_1,A_2},E_{A,B}$ are entanglement monotones that measure the internal and external entanglement, respectively. $E^{\max}_{A_1,A_2}$ is the maximum of $E_{A_1,A_2}$. This inequality is a general statement applicable to arbitrary quantum systems. It is possible to derive a more concise form for specific systems.} We start by considering the simplest model for the setup in Fig. ~\ref{Fig1} with subsystems $A_1,A_2,$ and B each contain a single qubit. We call this a 2+1 system. Assuming the full system is initialized in a pure state (\corr{we maintain this assumption throughout the entire paper}), we want to establish a relation between the external entanglement $E_{A,B}$ and the maximum internal entanglement $E^{\max}_{A_1,A_2}$ that can be created within subsystem A through subsequent unitary evolutions. In this work, we refer to this relation as the generalized monogamy of entanglement.

Let the reduced density matrix for subsystem A be $\rho_{A}$ and that for B be $\rho_B$. According to the Schmidt decomposition, they share the same spectrum:
\begin{equation}
\rho_A = \sum_{k=1}^2 \lambda_k \vert \lambda_k \rangle \langle \lambda_k \vert, \quad \rho_B = \sum_{k=1}^2 \lambda_k \vert \beta_k \rangle \langle \beta_k \vert.
      \label{PureStateSD}
\end{equation}
For clarity, $\lambda_k$ will always denote the eigenvalues of the reduced density matrices of the subsystems throughout this article. The decomposition on Eq.~\eqref{PureStateSD} shows that although subsystem A is a four-level system, we can understand its entanglement behavior by using a reduced two-level description. In that sense, system AB can be effectively regarded as a two-qubit system. Consequently,  the entanglement between A and B can be quantified through concurrence which can be obtained via a direct evaluation of Eq.~\eqref{cur} as:
\begin{comment}
From this, it follows that computing the entanglement entropy is independent of whether we use $\rho_A$ or $\rho_B$ to compute, it also makes evident that the smallest subsystem is the one constraining the amount of entanglement that can be shared between the two subsystems. It's important to note that the results described above only follow for pure states.
\end{comment}
\begin{equation}
    C_{AB}= 2 \sqrt{\text{det}(\rho_A)} = 2\sqrt{\lambda_1 - \lambda_1^2}\,.
    \label{C_AB}
\end{equation}
In this specific case, the tangle $\tau_{AB}\equiv C_{AB}^2$ \cite{PhysRevA.61.052306} is also equal to the linear entropy defined in Eq.~\eqref{l_ent}. 

To further investigate the maximum entanglement that can be created between $A_1$ and $A_2$, we use the result in Ref.~\cite{verstraete2001maximally} which states that for a generic two-qubit system represented by density matrix $\rho$ with eigenvalues ${\tilde{\lambda}_1,\tilde{\lambda}_2,\tilde{\lambda}_3,\tilde{\lambda}_4}$ in descending order, the maximum allowed entanglement can be characterized by the maximum concurrence as:
\begin{equation}
C_{\text{max}}=\max\left(0,\tilde{\lambda}_1-\tilde{\lambda}_3-2\sqrt{\tilde{\lambda}_2\tilde{\lambda}_4}\right) \,.
\label{CMAXQ}
\end{equation}
In our case, the density matrix we are interested in is $\rho_A$ and since $\lambda_3=\lambda_4=0$ according to Eq.~\eqref{PureStateSD}, the above equation reduces to: 
\begin{equation}
    C_{A_1A_2}^{max}=\lambda_1
    \label{C_12_max}\,,
\end{equation}
where we have assumed $\lambda_1>\lambda_2$. Combining Eqs.~\eqref{C_AB} and \eqref{C_12_max}:
\begin{equation}
    \left(C_{AB}\right)^2=4C_{A_1A_2}^{max}-4(C_{A_1A_2}^{max})^2 \,.
\end{equation}
Given that $\rho_A$ has a trace equal to one, it follows that $ 1\geq\lambda_1\geq1/2$. As a result, the above relation is monotonic and one-to-one, allowing to obtain an inverse relation:
\begin{equation}
    C_{A_1A_2}^{max}=\frac{1}{2}\left[1+\sqrt{1-(C_{A,B})^2}\right]\,.
    \label{CA1A2M}
\end{equation}

\begin{figure}[t!]
\includegraphics[width=0.48\textwidth]{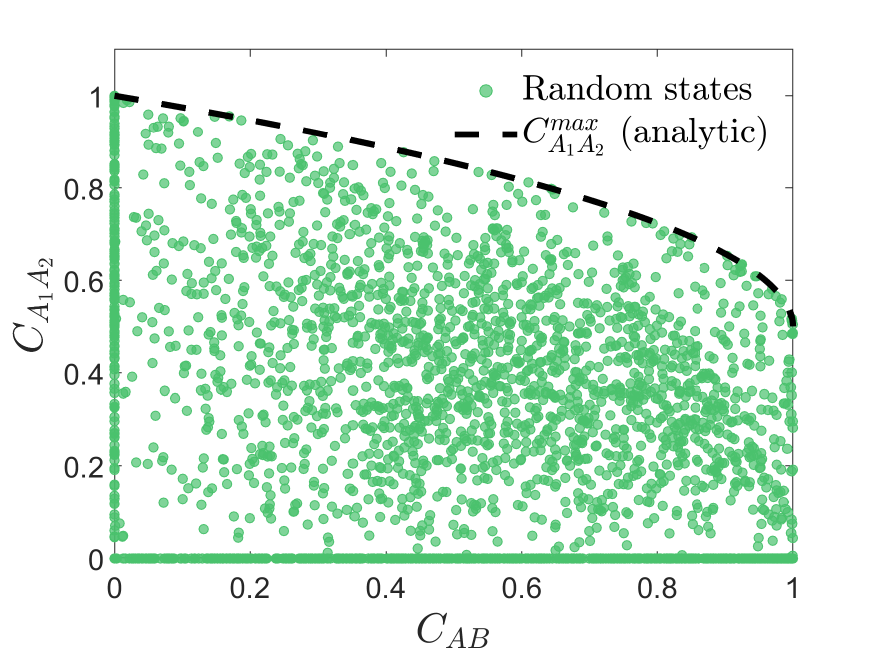}
\caption{$C_{A_1A_2}$ as a function of $C_{AB}$ for a set of 3000 randomly sampled three-qubit pure states. The dashed line represents the analytic expression for the maximum $C_{A_1A_2}$ allowed (Eq.~\eqref{CA1A2M}).}
\label{FigCMax}
\end{figure}

 In Fig.~\ref{FigCMax}, the relation between $C_{A_1A_2}$ and $C_{AB}$ is shown for a set of randomly sampled three-qubit pure states. The dashed line represents Eq.~\eqref{CA1A2M} and it is very clear from the figure that this expression is indeed the maximum value of $C_{A_1A_2}$ that can be obtained for a given $C_{AB}$. This monotonic relation is used as a motivation for our protocol proposed in the next section.

A similar relation as Eq.~\eqref{CA1A2M} for a pure three-qubit state involving negativity, instead of concurrence, can be derived. Explicitly writing out the partial transpose of $\rho_{AB}$ and solving for the eigenvalues, we obtain the normalized negativity (see \hyperref[AppA]{Appendix A}) as:
\begin{equation}
N_{AB} = 2\sqrt{\lambda_1(1-\lambda_1)}\,,
\label{N_AB_2}
\end{equation}
where the factor of two ensures normalization. In Ref.~\cite{verstraete2001maximally}, authors presented that, for a general two-qubit system described by a density matrix $\rho$ with eigenvalues ${\tilde{\lambda}_1,\tilde{\lambda}_2,\tilde{\lambda}_3,\tilde{\lambda}_4}$ in descending order, the maximum normalized negativity is given by:
\begin{equation}
    N^{max}(\rho) = \text{max} (0, \sqrt{(\tilde{\lambda}_1-\tilde{\lambda}_3)^2+(\tilde{\lambda}_2 - \tilde{\lambda}_4)^2}-\tilde{\lambda}_2 - \tilde{\lambda}_4) \,.
    \label{MaxN2Qubits}
\end{equation}
Again, with $\lambda_3=\lambda_4=0$, combining Eqs.~\eqref{N_AB_2} and \eqref{MaxN2Qubits} we find that the maximum negativity
within subsystem A can be written as a function of the negativity between A and B as:
\begin{equation}
N_{A_1A_2}^{max}=\frac{1}{2}\left(-1+\sqrt{4-2N_{AB}^2}+\sqrt{1-N_{AB}^2}\right)\,,
\label{MaxN2Monotonic}
\end{equation}
which, as expected, shows a monotonic one-to-one relation similar to Eq.~\eqref{CA1A2M}. \corr{We note that a similar derivation of this result and Eq.~\eqref{CA1A2M} can be found in Ref.~\cite{ExpMonogamy2020}}. 

To study larger systems, we increase the number of qubits in B while keeping A composed of two qubits $A_1$ and $A_2$. We shall examine the relations concerning negativities since the concurrences for such cases become hard to compute. Let subsystem B contain $N$ qubits with $N\geq 2$, the pure state of the full system can be described by:
\begin{equation}      |\psi\rangle=\sum_{i=1}^{4} {\sqrt{\lambda_i}\left|\lambda_i\right\rangle\otimes\left|\beta_i\right\rangle}\,.
\end{equation}
Now we have at most four non-zero eigenvalues of the respective reduced density matrices for A and B. The representation of the density matrix can be reduced to an effective $16 \times 16$ matrix independent of $N$, with six possible negative eigenvalues: $-\sqrt{\lambda_1 \lambda_2}$, $-\sqrt{\lambda_1 \lambda_3}$, $-\sqrt{\lambda_2 \lambda_3}$, $-\sqrt{\lambda_1 \lambda_4}$, $-\sqrt{\lambda_2 \lambda_4}$, $-\sqrt{\lambda_3 \lambda_4}$, and the negativity would be given by the sum of the absolute values of each of them. The maximum negativity following Eq.~\eqref{neg} is $N_0=3/2$ when all $\lambda_i$ are identical. Hence we define the normalized negativity for $AB$ as:
\begin{equation}
    N_{AB} = \frac{1}{3} \sum_{i,j=1}^4 (1-\delta_{ij}) \sqrt{\lambda_i \lambda_j}\,.
    \label{2+N}
\end{equation}
Note that we recover the case for $N=1$, up to a constant rescaling of $1/3$, when $\lambda_3=\lambda_4=0$, as expected. For this 2+$N$ case, $N^{max}_{A_1 A_2}$ is given by Eq.~\eqref{MaxN2Qubits} which, in general, cannot be further simplified. \corr{This causes the relation between $N^{max}_{A_1 A_2}$ and $N_{AB}$ to be described by a region rather than by a one-to-one curve. We discuss this more in detail in \hyperref[AppA]{Appendix A}. This general non-monotonic behavior is one of the challenges for our proposed protocol, as we will discuss further in the next section.}

\section{Bipartite entanglement quantification protocol}\label{SecIV}

To motivate our protocol for quantifying the entanglement between A and B, we first focus on the case of three qubits. We will use concurrence as an entanglement measure to explain the protocol, but negativity could be used as well, as we just showed that monotonic relations can be found for both in the 2+1 case.

\begin{figure}[t!]
\includegraphics[width=0.48\textwidth]{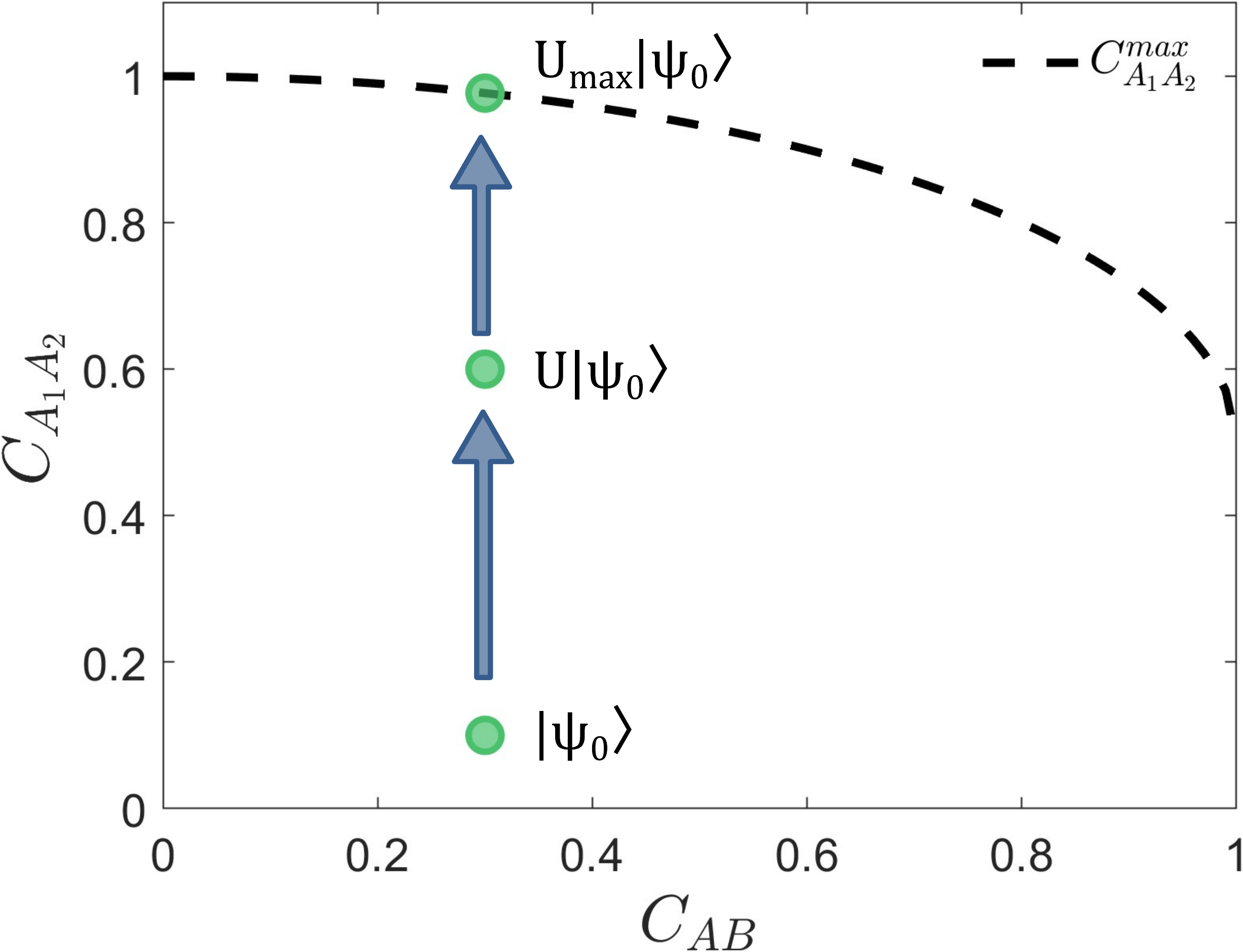}
\caption{Protocol for quantification of the entanglement between A and B. An initial state $\psi_0$ is evolved under a unitary transformation $U$ that acts only on subsystem A. An optimal unitary $U_{max}$ would move the state to the boundary of maximum $C_{A_1A_2}$. Measurement of $C_{A_1A_2}$ and the relation on Eq.~(\ref{CA1A2M}) are then used to quantify $C_{A,B}$.}
\label{FigProtocolSchematics}
\end{figure}

Let us start with a given pure state $\vert \psi_0 \rangle$ (see Fig.~\ref{FigProtocolSchematics}) with a given $C_{AB}$, which we want to quantify, and $C_{A_1A_2} \neq C_{A_1A_2}^{max}$. We then consider a unitary transformation $U$ that only acts on subsystem A and creates entanglement between $A_1$ and $A_2$. This will move the state upward in the diagram as $C_{A_1A_2}$ increases, but it will not move horizontally because, by definition, local unitary operations on A cannot change the bipartite entanglement between A and B. Suppose that through some optimization we find $U_{max}$ that pushes $C_{A_1A_2}$ to the boundary, we could then use Eq.~(\ref{CA1A2M}) and the measurement of $C_{A_1A_2}$ to quantify the entanglement between A and B ($C_{AB}$). \corr{As discussed in Ref.~\cite{Calamet2018}, for a pure state, such an optimal unitary transformation can always be found for a large class of entanglement monotones. As we show later, this is not necessarily the case when we consider, for example, squeezing parameters.}

The protocol just described is an ideal case. In light of the discussions in previous sections, when larger systems are considered, our protocol would require modifications. We identify two challenges that we need to address to generate useful insights with this approach. First, for our protocol to work, we need a way to quantify either $C_{A_1,A_2}$ or $N_{A_1,A_2}$. However, these quantities are hard to access, as we pointed out earlier. Spin-squeezing parameters could be an interesting candidate for characterizing entanglement in spin systems, as they involve only the measurement of total spin expectation values and their fluctuations. However, even if spin-squeezing parameters can be used to determine whether a system is entangled or not, they cannot generally be used to quantify how much entanglement is present in the system \cite{MA201189}.

Second, the fact that the relation between the entanglement of A and B and the maximal entanglement that can be generated within A is non-monotonic, and potentially widely spread, will damage the predictive power of the protocol. \corr{However, we note that this is the expected behavior for generic random states}. If the states we want to characterize are generated from a known family of initial states and then evolved with a known family of Hamiltonians with specific symmetries and conserved quantities, which is often the case in practical situations, then we would expect the relation between the two quantities of interest to become simpler, though it could still be non-monotonic.

In the next couple of sections, we address both of these challenges and show examples on how our protocol can be used effectively.

\section{Spin-squeezing and entanglement}\label{SecV}

Spin-squeezed states are great assets in the field of quantum metrology and have been recently realized in a wide range of platforms \cite{leroux2010,chen2011,Braverman2019,bornet2023scalable,eckner2023realizing,Hines2023}. Spin-squeezing parameters are strongly connected to the entanglement measures introduced in the previous sections. For instance, in Ref.~\cite{Toth2009} a complete set of spin-squeezing inequalities was provided, and the violation of them signifies a sufficient condition for non-separability. We should note, however, that most bipartite entanglement criteria defined in terms of spin-squeezing parameters are expected to determine whether a state is entangled or not (related to the PPT criterion) but do not necessarily map the spin-squeezing parameters into other entanglement measures such as concurrence, negativity, etc. This means that most of these criteria cannot necessarily quantify entanglement properly, except for some special cases \cite{wang2002pairwise}. 

We investigate the relationship between squeezing parameters and linear entropy for different types of states in the following sections. To carry out the analysis, we start by introducing the squeezing operator we will be considering here:
\begin{equation}
\xi^2 = \frac{4 \text{ min}((\Delta J_{\vec{n}_\perp})^2)}{N}\,,
    \label{kitagawauedaparam}
\end{equation}
where $N$ is the number of spin 1/2 particles considered, and $J_{\vec{n}_\perp}$ is the spin angular momentum operator in the direction $\vec{n}_{\perp}$, which is any direction perpendicular to the mean spin direction $\vec{n} = \langle \vec{J} \rangle / \vert \langle \vec{J} \rangle \vert$, with $\langle \vec{J} \rangle = (\langle J_x\rangle, \langle J_y \rangle, \langle J_z\rangle)^T$. $\Delta O = \sqrt{\langle O^2 \rangle - \langle O \rangle^2}$ denotes the uncertainty of an operator $O$, and the minimization is carried out with respect to all perpendicular directions $\vec{n}_\perp$. This spin-squeezing operator was first introduced by Kitagawa and Ueda \cite{Kitagawa1} and it is the one we will be using throughout the subsequent analysis, for a review of other spin-squeezing operators we refer the reader to Ref.~\cite{MA201189}.

\section{Example Spin Systems} \label{SecVI}
In this section, we test the effectiveness of our protocol on spin $1/2$ systems of different sizes. Each one consists of two subsystems A and B with an equal number of qubits. For all cases, we consider the system to be initialized in a pure state with all qubits in $\ket{\downarrow}$ state. We first apply a Hamiltonian $H_{AB}$ that creates entanglement between A and B. At each evolution time $t$ we characterize the entanglement between A and B using the linear entropy in Eq.~\eqref{l_ent}, denoted as $S_{L,AB}$. Subsequently, we apply a local Hamiltonian $H_A$ on system A to maximize the spin squeezing within it such that the squeezing parameter $\xi^2_{A}$ becomes minimal. We study the relation between the above parameters and show that $\text{min}(\xi^2_{A})$ can be a good predictor of $S_{L,AB}$ when the appropriate $H_A$ is chosen.

In principle, the choice of $H_{AB}$ can be arbitrary, and the determination of the optimal $H_A$, which will depend on the specific generated state, can be hard to calculate. Here we focus on some common types of Hamiltonians that are used to generate spin squeezing and entanglement, including one-axis twisting (OAT), one-axis twisting with transverse field (TF), and two-axis twisting (TAT) \cite{MA201189} Hamiltonians which are explicitly defined as:
\begin{eqnarray}
    &&H_{OAT} = \Omega J_x^2, \; \nonumber \\ &&  H_{TF} = \Omega J_x^2 + \omega J_z \,,\nonumber \\
    && H_{TAT} = \Omega (J_x J_y + J_y J_x)\,.
    \label{H_list}
\end{eqnarray}
For simplicity, we will set $\omega=\Omega$ in $H_{TF}$ throughout this paper. To obtain some analytical results, we also consider a GHZ Hamiltonian that can create an $N$-body GHZ state: 
\begin{equation}
H_{\text{GHZ}} = \Omega \prod_i \sigma_i^x,
\label{H_ghz}
\end{equation}
where $\sigma_i^x$ is the Pauli-$x$ operator on spin $i$.  
\\

\begin{figure}[t!]
    \centering
    \includegraphics[width=0.45\textwidth]{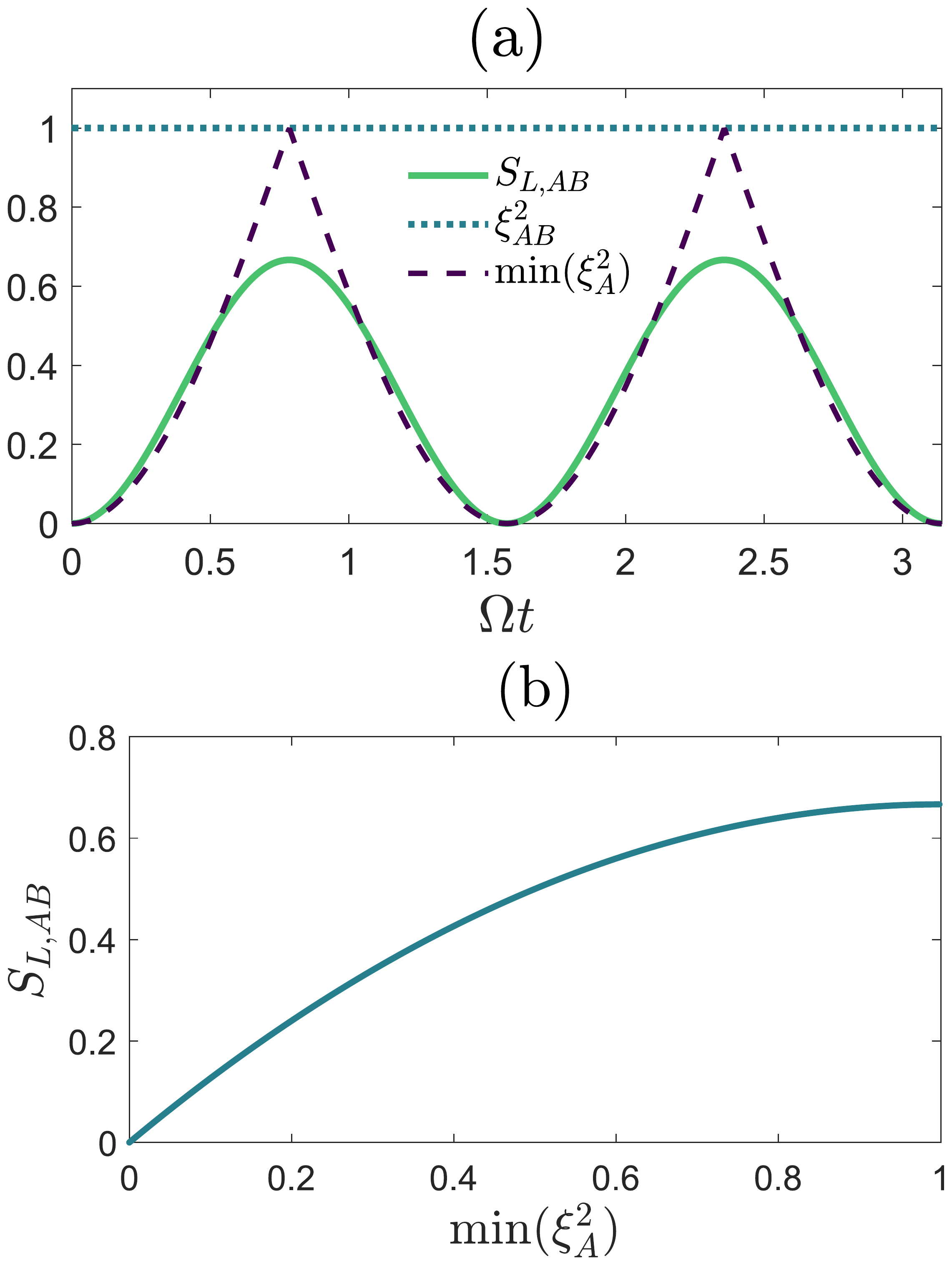}
    \caption{Plots for a four qubit system with $H_{AB}$ and $H_A$ being a four-body and two body GHZ Hamiltonian, respectively. (a) Linear entropy $S_{L,AB}$ and spin squeezing parameters $\xi_{AB}^2$ and $\text{min}(\xi_{A}^2)$ as a function of $t$. (b) Linear entropy $S_{L,AB}$ as a function of the minimum spin squeezing parameter $\text{min}(\xi_{A}^2)$ in subsystem A.}
    \label{fig:simple_H}
\end{figure}

\subsection{Four qubits with GHZ Hamiltonian}
To develop an intuitive understanding, we begin by examining a system composed of four qubits where qubits 1 and 2 constitute subsystem A, while qubits 3 and 4 constitute subsystem B. First, we evolve the system with the GHZ Hamiltonian $H_{AB}=H_{\text{GHZ}}$ defined in Eq.~\eqref{H_ghz} to create entanglement between A and B. We note that the subspace spanned by $\{\ket{\downarrow\downarrow \downarrow \downarrow}\equiv\ket{e_0} ,\ket{\uparrow \uparrow \uparrow \uparrow}\equiv\ket{e_1}\}$ is invariant under $H_{AB}$, and since the initial state is taken to be $\vert \psi_0 \rangle =\vert e_0 \rangle$, we remain in this subspace throughout the evolution. Following the results in previous sections, the states generated under this Hamiltonian should display a monotonic relationship between the bipartite entanglement of A and B and the maximum entanglement that can be generated in A (all states in the dynamics are states with $\lambda_3=\lambda_4=0$, see purple circles and red dashed line in Fig.~\ref{Nmax} in \hyperref[AppA]{Appendix A}).
The density matrix at time $t$, in the $\{ \ket{e_0},\ket{e_1}\}$ basis, is given by:
\begin{equation}
 \rho_{AB}(t)=\frac{1}{2}\left[\begin{matrix}1+\cos{\left(2\phi\right)}&i\sin{(2\phi})\\-i\sin{(2\phi})&1-\cos{\left(2\phi\right)}\ \\\end{matrix}\right],\ \ \phi=\Omega t\,.
\end{equation}
Using the definition in Eq.~\eqref{kitagawauedaparam}, the spin-squeezing parameter is found to be a constant $
    \xi_{AB}^2 = 1$ independent of time.
The reduced density matrix for subsystem A in the basis $\{ \ket{\downarrow \downarrow},\ket{\uparrow \uparrow}\}$ takes the form:
\begin{equation}
    \rho_A(t)=\frac{1}{2}\left[\begin{matrix}1+\cos{\left(2\phi\right)}&0\\0&1-\cos{\left(2\phi\right)}\ \\\end{matrix}\right]\,.
    \label{red_A}
\end{equation}
It follows that we can characterize the entanglement between subsystems A and B using linear entropy as defined in Eq.~\eqref{l_ent}:
\begin{equation}
    S_{L,AB}  =  \frac{2}{3}[1-\cos^2{\left(2\phi\right)}]\,.
    \label{mix}
\end{equation}

Now that we can fully characterize the system at time $t$, we consider the subsequent application of a local GHZ Hamiltonian $ H_{A} = \Omega' \sigma_1^x \sigma_2^x$ on subsystem A for some time $t'$. The spin-squeezing parameter for subsystem A is given by:
\begin{equation}
    \xi_{A}^2= 1-\left|\cos{2\phi}\sin{2\phi^\prime}\right|,\; \;\phi'=\Omega't'\,.
\end{equation}
Minimizing $\xi_{A}^2$ with respect to $\phi'$ to find the optimal evolution time $t'$, gives the maximum spin squeezing allowed in subsystem A:
\begin{equation}
    \text{min}(\xi_{A}^2)= 1-\left|\cos{2\phi}\right| \,.
\end{equation}
Plots of $S_{AB}$, $\xi_{AB}^2$, and  $\text{min}(\xi_{A}^2)$ as functions of $t$ are shown in Fig.~\ref{fig:simple_H}(a). The linear entropy in Eq.~\eqref{mix} can be then rewritten in the form:
\begin{equation}
    S_{L,AB} = \frac{2}{3}\left[1-\left(1-\text{min}(\xi_{A}^2)\right)^2\right]\,,
    \label{GHZLin}
\end{equation}
which is monotonic for $\text{min}(\xi_{A}^2)\in[0,1]$ as illustrated in Fig.~\ref{fig:simple_H}(b). In this simple example, we showed that the total spin-squeezing parameter of the system is time-independent, failing to predict the bipartite entanglement between A and B. However, the minimum squeezing parameter for A shows a monotonic relation with the linear entropy, illustrating a clear usage of our protocol. For instance, if we are interested in characterizing a state resulting from the evolution under $H_{AB}$ for a time $t$, and we can repeat this state preparation with high accuracy, we could estimate the minimum value of $\xi_{A}^2$ by measuring subsystem A for different evolution times $t'$. $\text{min}(\xi_{A}^2)$ can then be used to determine $S_{L, AB}$ through Eq.~\eqref{GHZLin}. This is how our protocol exploits the property of the generalized monogamy of entanglement we proposed earlier to quantify the entanglement between A and B.

In terms of efficiency, our protocol still requires the repeated preparation of the state we want to characterize, just as for quantum state tomography. Nonetheless, the number of realizations might be reduced by an efficient choice of the different values of $t'$. This could speed up the protocol considerably as long as the profile of $\xi_{A}^2(t)$ has some periodicity and not too many local minima. 

\subsection{Four and eight qubit systems with $H_{AB}=H_{OAT}$}

\begin{figure}[t!]
    \includegraphics[width=0.45\textwidth]{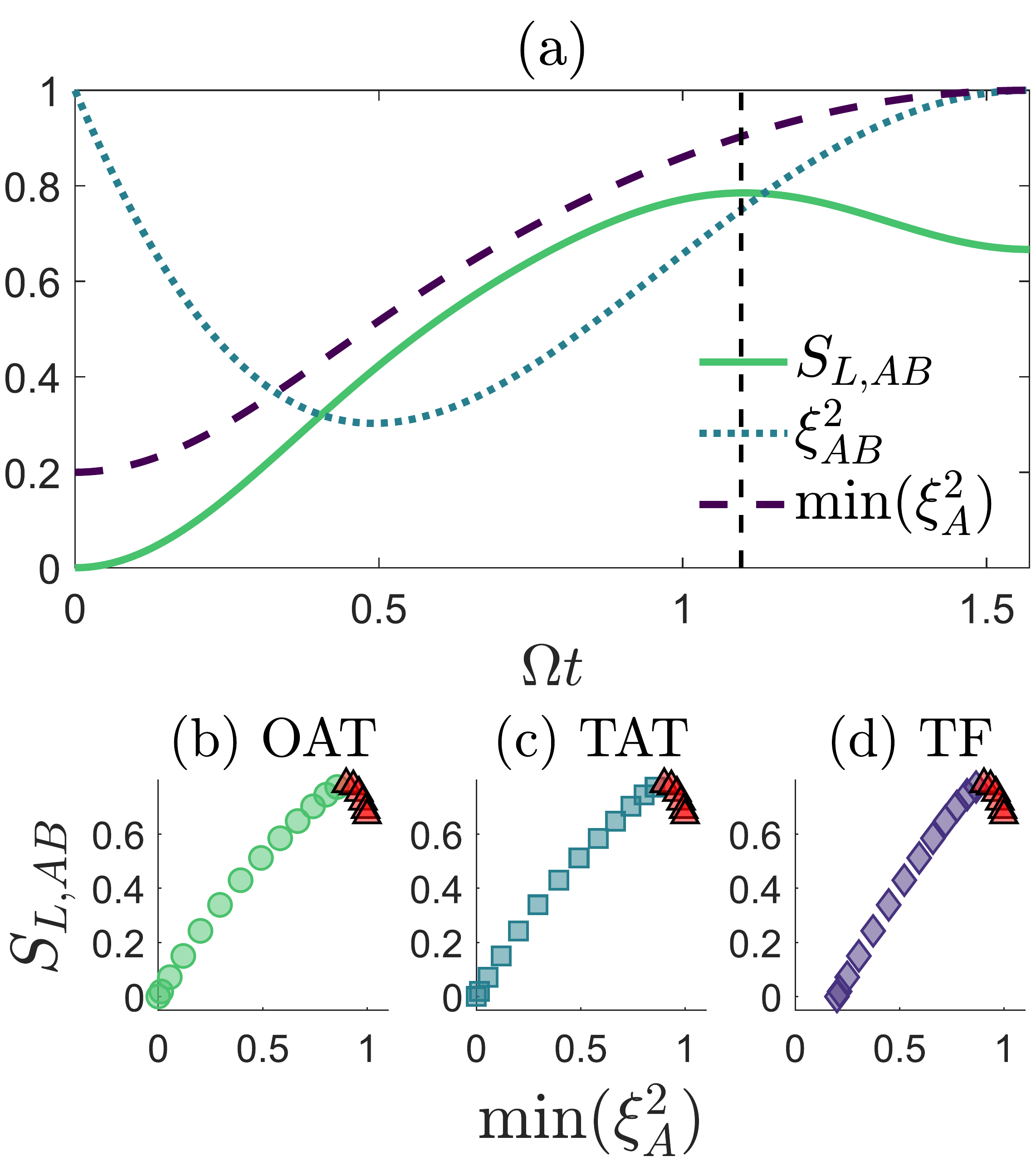}
    \caption{Four-qubit system with $H_{AB}=H_{OAT}$. (a) Linear entropy $S_{L,AB}$ and spin squeezing parameters $\xi_{AB}^2$ and $\text{min}(\xi_{A}^2)$ as a function of time with $H_A=H_{TF}$. \corr{The vertical dashed line signals the} section of the dynamics where the one-to-one relation between $\text{min}(\xi_{A}^2)$ and $S_{L,AB}$ is broken. \corr{Data points corresponding to this section are denoted by red triangles in the subsequent panels}. The linear entropy $S_{L,AB}$ as a function of minimum spin squeezing parameter $\text{min}(\xi_{A}^2)$ in subsystem A is shown for (b) $H_A=H_{OAT}$, (c) $H_A=H_{TAT}$, and (d) $H_A=H_{TF}$.}
    \label{fig:4q_OAT}
\end{figure}

We now consider a more experimentally relevant case where $H_{AB} = H_{OAT}$. We optimize the spin squeezing in subsystem A with the three Hamiltonians in Eq.~\eqref{H_list}. The subspace spanned by $\{ \ket{e_0},\ket{e_1}\}$ is no longer preserved by $H_{AB}$, so we study the dynamics numerically. The results are shown in Fig.~\ref{fig:4q_OAT} for a total system of four qubits (two qubits in A and two in B). Since the evolution under $H_{OAT}$ is periodic we only show the evolution of half a period, that is $0\leq \Omega t \leq \pi/2$. For all choices of $H_A$, below $S_{L,AB} \sim 0.6$ there exists a one-to-one relation between $\text{min}(\xi_{A}^2)$ and $S_{L,AB}$. The similar performances of these Hamiltonians are consistent with the fact that for a subsystem of two qubits, all these Hamiltonians describe very similar dynamics. 

The section of the time evolution that breaks the one-to-one relation is highlighted with black in Fig.~\ref{fig:4q_OAT}. In this region $\text{min}(\xi_{A}^2)$ continues to monotonically increase, while $S_{L,AB}$ develops a more complex behavior, which results in reduced accuracy of the protocol for this region. An ideal scenario would be that of the GHZ Hamiltonian case in Fig.~\ref{fig:simple_H}(a) where both parameters follow the same trends and then the protocol is highly accurate for all ranges of $S_{L,AB}$.  To demonstrate why the protocol fails in this region, we notice that the state at $\Omega t = \pi/2$, where $\text{min}(\xi_{A}^2)$ becomes maximum but $S_{L,AB}$ reaches a local minimum, is exactly a 4-body GHZ state. Hence the reduced density matrix of subsystem A is a statistical mixed ensemble of states $\ket{\downarrow \downarrow}$ and $\ket{\uparrow \uparrow}$ with equal weights. Due to the special symmetry of a two-qubit system ($J_x^2=\sigma_1^x\sigma_2^x/4$), this state remains invariant under all three $H_A$ we chose and no squeezing can be generated using these Hamiltonians ($\text{min}(\xi_{A}^2)=1$).  Here, no further squeezing can be generated with these specific Hamiltonians in Eq.~\eqref{H_list}, however, there might exist some local unitary that further generates squeezing in A for this state \cite{verstraete2001maximally}, but it can be very difficult to generate the corresponding Hamiltonian. 

\begin{figure}[t!]
    \centering
    \includegraphics[width=0.45\textwidth]{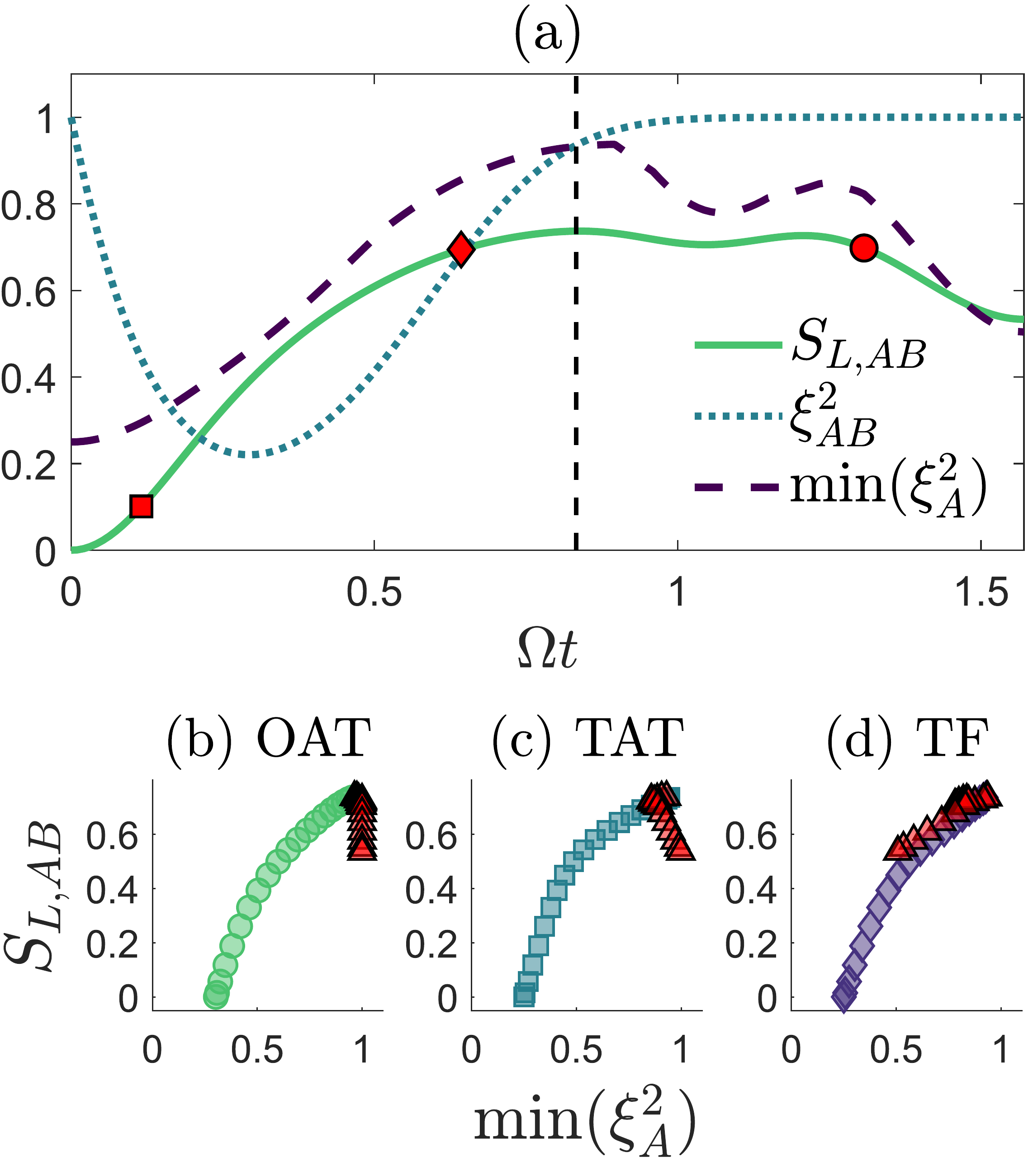}
    \caption{Eight-qubit system with $H_{AB}=H_{OAT}$. (a) Linear entropy $S_{L,AB}$ and spin squeezing parameters $\xi_{AB}^2$ and $\text{min}(\xi_{A}^2)$ as a function of time with $H_A=H_{TF}$. Three specific states signaled with square, diamond, and circle markers are selected for further exploration in Fig.~\ref{fig:dH}. We denote them as $p_1$, $p_2$, and $p_3$, respectively. \corr{The vertical dashed line signals the} section of the dynamics where the one-to-one relation between $\text{min}(\xi_{A}^2)$ and $S_{L,AB}$ is broken. \corr{Data points corresponding to this section are denoted by red triangles in the subsequent panels}. The linear entropy $S_{L,AB}$ as a function of minimum spin squeezing parameter $\text{min}(\xi_{A}^2)$ in subsystem A is shown for (b) $H_A=H_{OAT}$, (c) $H_A=H_{TAT}$, and (d) $H_A=H_{TF}$.}
    \label{fig:8q_OAT}
\end{figure}

In more generic circumstances, especially when subsystem A contains more than 2 qubits, one of the three Hamiltonians $H_{A}$ might outperform the others. We illustrate this by applying our protocol in a system of a total of eight qubits with each subsystem containing four of them, the results are presented in Fig.~\ref{fig:8q_OAT}.

The first thing we notice in Fig.~\ref{fig:8q_OAT}(a) is that again the linear entropy has a non-monotonic behavior in each period of the evolution. However, now $\text{min}(\xi_{A}^2)$ develops very similar features when subsystem A is evolved under $H_A=H_{TF}$. To visualize this we plot the relation between those two variables in Fig.~\ref{fig:8q_OAT}(d), where we see a nearly monotonic relation, meaning the optimal squeezing parameter of A would be a good predictor of the bipartite entanglement of A and B. By contrast, if $H_A$ is chosen to be the one- or two-axis twisting Hamiltonian, the relation deviates far from a monotonic behavior for highly entangled states (panels (b) and (c)). Additionally, we point out that for both four and eight total qubits, the squeezing parameter of the total system ($\xi_{AB}^2$) is a bad predictor of the linear entropy for that state.

\subsection{Effects of different $H_A$}
The better performance of $H_A = H_{TF}$ over the other two options can be understood as follows. Based on the earlier sections, the entanglement between A and B limits the maximum entanglement that can be generated within A. Then, the ideal scenario for the proposed protocol is a Hamiltonian $H_A$ for which the maximum entanglement within A is reached exactly at the point with the minimal squeezing parameter. 
 
 Let us momentarily focus on subsystem A. We consider three different states generated under $H_{AB}=H_{OAT}$ denoted by red markers in Fig.~\ref{fig:8q_OAT}(a). To obtain the value of $\text{min}\left(\xi_{A}^2\right)$ we evolve each state with the three possible Hamiltonians $H_A$ in Eq.~\eqref{H_list} for a long enough time to make sure we explore enough states in the available Hilbert space ($\Omega t =100$). For each point of this evolution, we compute the spin squeezing $\xi_{A}^2$ and entanglement negativity $N_{A}$. Note that since the state of A is generally mixed (given non-zero bipartite entanglement between A and B), we cannot use linear entropy as a measure of entanglement within subsystem A. 

\begin{figure}[t!]
    \includegraphics[width=0.45\textwidth]{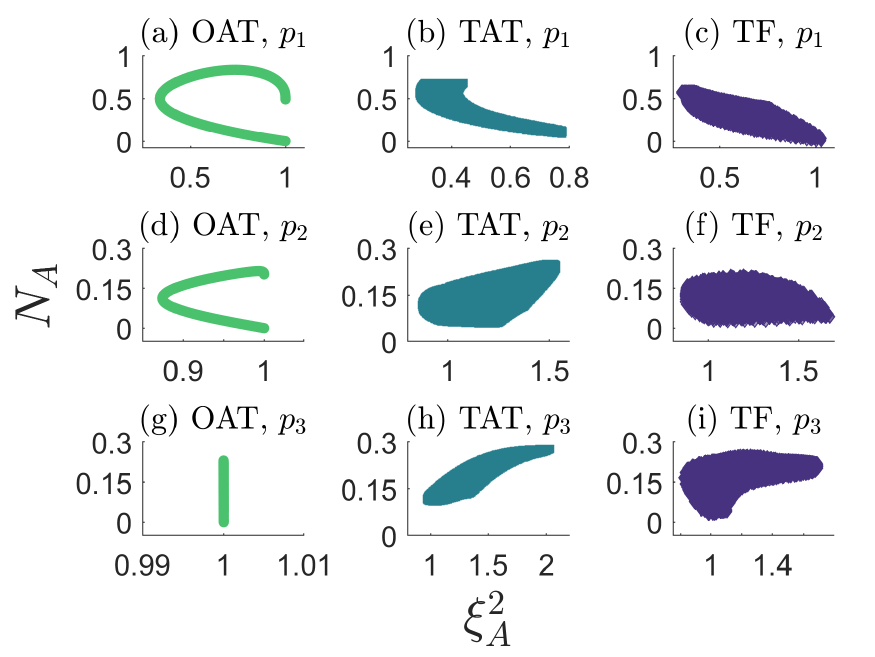}
    \caption{Effects of different $H_A$ on capturing maximum entanglement. Three states with different $S_{L,AB}$ are selected from the dynamics under $H_{AB}=H_{OAT}$ and then evolved with three different Hamiltonians $H_A$. The negativity of subsystem A is plotted as a function of the total spin squeezing of that subsystem. From left to right the Hamiltonian $H_A$ which evolves subsystem A is given by OAT, TAT, and TF, respectively. From top to bottom, the initial mixed state corresponds to $p_1$, $p_2$, and $p_3$, respectively (see Fig.~\ref{fig:8q_OAT}).}
    \label{fig:dH}
\end{figure}

Figure \ref{fig:dH} shows the results for states $p_1$, $p_2$, and $p_3$, which correspond to linear entropies $S_{L,AB}=0.1$, $0.7$, and $0.7$, respectively. The difference between $p_2$ and $p_3$ is that the latter is a state in the non-monotonic regime (see black solid lines in Fig.~\ref{fig:8q_OAT}(a)). If we use as a criterion of optimality whether the minimum squeezing is close to the value of maximum entanglement (maximum negativity here), the Hamiltonian $H_{TF}$ is more optimal than the other two for all cases. For $p_3$, all Hamiltonians perform worse than for $p_2$ and $p_1$ which is consistent with the results in Fig.~\ref{fig:8q_OAT}(b), (c), and (d). 

We note that the range of values of squeezing that $H_A=H_{OAT}$ can generate, considerably shrinks as the bipartite entanglement $S_{L,AB}$ grows. For state $p_3$, almost no squeezing can be generated. $H_A=H_{TAT}$, on the other hand, starts to anti-squeeze the spin-ensemble A, resulting in the squeezing parameter shifting to larger values. Under the evolution with $H_A=H_{TF}$, the general relation between $N_A$ and $\xi_{A}^2$ is more complex than for the other two. However, it is clear that the smallest spin-squeezing always corresponds to a value of $N_A$ that is not too far from the maximum possible value, making it the optimal choice for $H_A$ in our protocol. 

We emphasize that the latter results are valid for states generated under $H_{AB}=H_{OAT}$. In general, the definition of an optimal Hamiltonian $H_A$ is related to the symmetries of both $H_A$ and $H_{AB}$. As we discussed for $\Omega t= \pi/2$ in Fig.~\ref{fig:4q_OAT}(a), the dynamics might lead to highly symmetric mixed states in A. To squeeze these states further, the breaking of some of those symmetries is required to explore a larger region of the Hilbert space where highly squeezed states can be accessed under the dynamics. For the Hamiltonians in Eq.~\eqref{H_list}, both $H_{OAT}$ and $H_{TAT}$ are symmetric under the change $\vec{J} \rightarrow -\vec{J}$. Additionally, the one-axis twisting Hamiltonian is invariant under rotations around the $x$-axis while the two-axis twisting Hamiltonian has a discrete symmetry (parity) related to rotations around the $z$-axis, namely, $J_x \rightarrow -J_x$ and $J_y \rightarrow -J_y$. All of the above symmetries are not preserved in $H_{TF}$, leading to the intuition that an optimal $H_A$ would be a very random-like Hamiltonian where no symmetries are conserved. Nonetheless, the optimality criterion of the minimal squeezing parameter being achieved near the region of maximum entanglement might not be fulfilled by any generic Hamiltonian, making the determination of an optimal $H_A$ non-trivial. For this work, however, we focused on a small family of Hamiltonians that have metrological importance and can be realized in current state-of-the-art experimental platforms, making it easier to conclude that $H_{TF}$ is the optimal choice for $H_A$.

\corr{\subsection{Larger $N$}}

\begin{figure}[t!]
    \includegraphics[width=0.45\textwidth]{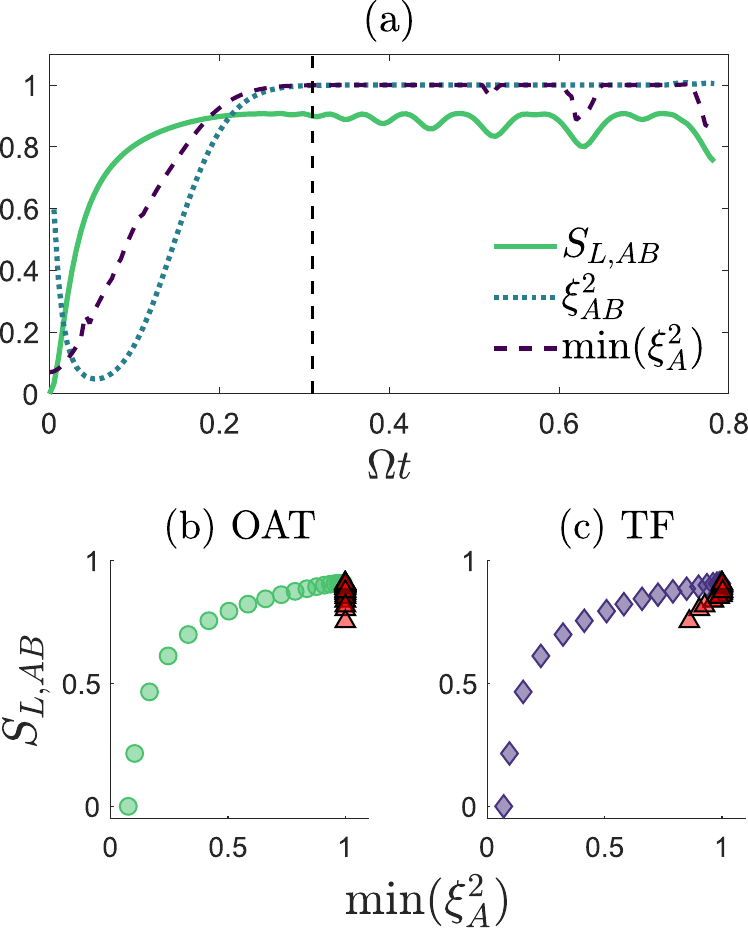}
    \caption{\corr{One hundred-qubit system with $H_{AB}=H_{OAT}$. Linear entropy $S_{L,AB}$ and spin squeezing parameters $\xi_{AB}^2$ and $\text{min}(\xi_{A}^2)$ as a function of time with $H_A=H_{TF}$. The vertical dashed line signals the section of the dynamics where the one-to-one relation between $\text{min}(\xi_{A}^2)$ and $S_{L,AB}$ is broken. Data points corresponding to this section are denoted by red triangles in the subsequent panels. The linear entropy $S_{L,AB}$ as a function of minimum spin squeezing parameter $\text{min}(\xi_{A}^2)$ in subsystem A is shown for (b) $H_A=H_{OAT}$  and (c) $H_A=H_{TF}$.}}
    \label{fig:100:q_OAT}
\end{figure}

\corr{Finally, we explore if the results obtained above can be extended to systems with a larger number of qubits. This is especially important since quantum state tomography become more complicated to perform in larger systems. In Fig.~\ref{fig:100:q_OAT}, we show the results of the protocol for $N=100$ qubits with 50 qubits in A and 50 qubits in B. Details of the numerical calculations for larger $N$ can be found in \cite{code_repository}.}

\corr{We confirm that the trends presented earlier remain true for larger systems. For times shorter than those denoted by a vertical dashed line in panel (a) of Fig.~\ref{fig:100:q_OAT} (where the one-to-one relationship is broken) both $H_A=H_{OAT}$ and $H_A = H_{TF}$ perform optimally, while for later times $H_{TF}$ performs significantly better.  We note that for this case we only take the evolution of the total system for times $0\leq \Omega t \leq \pi/4$. For times $\pi/4\leq \Omega t \leq \pi/2$, the protocol performs worse, indicating that the range of times when the protocol performs better decreases with $N$. However, we note that for the time period considered in Fig.~\ref{fig:100:q_OAT}, a large variety of entangled states can still be characterized appropriately by our protocol. For instance, a lot of experiments where the total system is evolved with $H_{AB} = H_{OAT}$ would be operated in times around the optimal squeezing time (minimum of $\xi_{AB}^2$ in Fig.~\ref{fig:100:q_OAT}(a) at around $\Omega t \approx 0.05$) which lie well within the region of optimal operation of our protocol.} Of course, this is still a modest number of qubits, and an experimental implementation of the protocol described here would be the best way to confirm these trends for systems of thousands of qubits, or larger. 
\vspace{1.5\baselineskip}
\section{Conclusions} \label{SecVII}

In conclusion, we have \corr{used the concept of generalized monogamy of entanglement} to conceptually introduce a protocol to quantify the bipartite entanglement \corr{in a system of pure state} by a subsequent coherent evolution and measurement of \corr{spin squeezing on} one of the subsystems. \corr{The spin squeezing parameter can be obtained by measuring the fluctuation of the collective spin operator, which is generally a much easier measurement than the entanglement entropy.} We show that in some specific cases, such as in the GHZ Hamiltonian, the protocol works with near-perfect accuracy (which would only be limited by experimental errors), while in other more experimentally relevant cases, the accuracy of the protocol \corr{is very high for short times} and can be pushed by a judicious selection of the subsystem Hamiltonian $H_A$. \corr{We want to remark that the monogamy inequality Eq.~(\ref{E_ineq}) relies on the convex measure for a quantum resource. Spin squeezing is a sufficient condition for entanglement. However, the convexity of spin squeezing in this context has not been addressed and is worth investigating in the future.}

One advantage of our protocol, besides the challenging techniques required for efficient full quantum state tomography, is the fact that entangled subsystems (A and B here) can be separated far apart \cite{hofmann2012,bernien2013heralded}, especially for applications on quantum communication. In that case, only local operations (on a given subsystem) and communication with other parties are available to quantify the entanglement between subsystems, our protocol is then readily available to be integrated into these types of applications. \corr{Of course, in the case of an initial pure state, tomography on one of the subsystems is enough to characterize the bipartite entanglement, however, this is not the case for mixed systems, where we expect our protocol to still be applicable, as discussed below.}

Although we have only focused on spin systems due to the recent progress in the generation of spin-squeezed states in multiple platforms, the ideas presented here can be extended to heterogeneous systems where spins are coupled to a different set of degrees of freedom. A notion of how the entanglement within a subsystem is limited by the bipartite entanglement between subsystems in such setups could complement, for example, the results on recently proposed protocols to generate spin-squeezed states through the oscillation of squeezing between the spin and bosonic Hilbert spaces \cite{Barberena_2024,RLS2024}. We also emphasize that the results presented here are restricted to the case where the total system composed of A and B is in a pure state. In general, mixing of the state would decrease the accuracy of the protocol as now the squeezing of subsystem A can be influenced not only by the bipartite entanglement between A and B but also by the classical correlations of the classical mixture. Nonetheless, given the cleanliness and high control in current experimental platforms, we expect that target entangled states can be generated with a sufficiently low degree of mixing and, consequently, our protocol could potentially work to high accuracy in such cases although further numerical exploration to confirm this is pending.

Finally, we would like to point out that a worth-exploring direction would be to analyze \corr{this protocol in terms of other entanglement witnesses like} mutual information, which characterizes the total amount of correlations between two subsystems \cite{Groisman2005}. This quantity has been used to investigate area law entanglement \cite{Wolf2008}, and obtaining some results in this language would allow connections to previous results in entanglement rate bounds \cite{VAcoleyen2013} and correlation length decay \cite{brandao2013area}. \corr{More related to the protocol presented here, in Ref.~\cite{ExpMonogamy2021}, the generalized monogamy of entanglement relations were probed experimentally using the quantum mutual information of a pair of photons.}

\begin{acknowledgements}
The authors acknowledge Prof. Shuming Cheng and Prof. Zhexuan Gong for insightful discussions. H.P.
acknowledges support from the US NSF and the Welch Foundation (Grant No. C-1669).

\end{acknowledgements}

\bibliography{refs}{}

\section{Appendix A: Negativity for 2+N systems}\label{AppA}

For a three-qubit system (two qubits on A and one on B) on a pure state described by Eq.\eqref{PureStateSD}, the density matrix of the composite system can be effectively represented as a four-dimensional matrix on the basis of $\ket{\lambda_i}\otimes\ket{\beta_j},\;i,j=1,2$:
\begin{equation}
    \rho_{AB}=\left[\begin{matrix}\lambda_1&0&0&\sqrt{\lambda_1(1-\lambda_1)}\\0&0&0&0\\0&0&0&0\\\sqrt{\lambda_1(1-\lambda_1)}&0&0&\lambda_1\\\end{matrix}\right]\,.  
\end{equation}
Taking partial transpose with respect to subsystem A gives:
\begin{equation}
    \rho^{T_A}_{AB}=\left[\begin{matrix}\lambda_1&0&0&0\\0&0&\sqrt{\lambda_1(1-\lambda_1)}&0\\0&\sqrt{\lambda_1(1-\lambda_1)}&0&0\\0&0&0&1-\lambda_1\\\end{matrix}\right]\,.
\end{equation}
By explicitly diagonalizing $\rho^{T_A}_{AB}$ we can identify $\lambda_4^{T_A} =-\sqrt{\lambda_1(1-\lambda_1)} $ as the only possible negative eigenvalue. Substituting this result into the definition of normalized negativity gives us Eq.~\eqref{N_AB_2}. 

Following a similar procedure we can derive the negativity for the system of 2+N qubits. In this case, $\rho^{T_A}_{AB}$ is given by a  $16 \times 16$ dimensional matrix that contains six possible negative eigenvalues following the form $(\delta_{ij}-1)\sqrt{\lambda_i \lambda_j}$, where $i,j=1,2,3,4$, and $\sum_{i=1}^4\lambda_i =1$. Consequently, the normalized negativity for this case is given by Eq.~\eqref{2+N}.  

\begin{figure}[t!]
\includegraphics[width=0.49\textwidth]{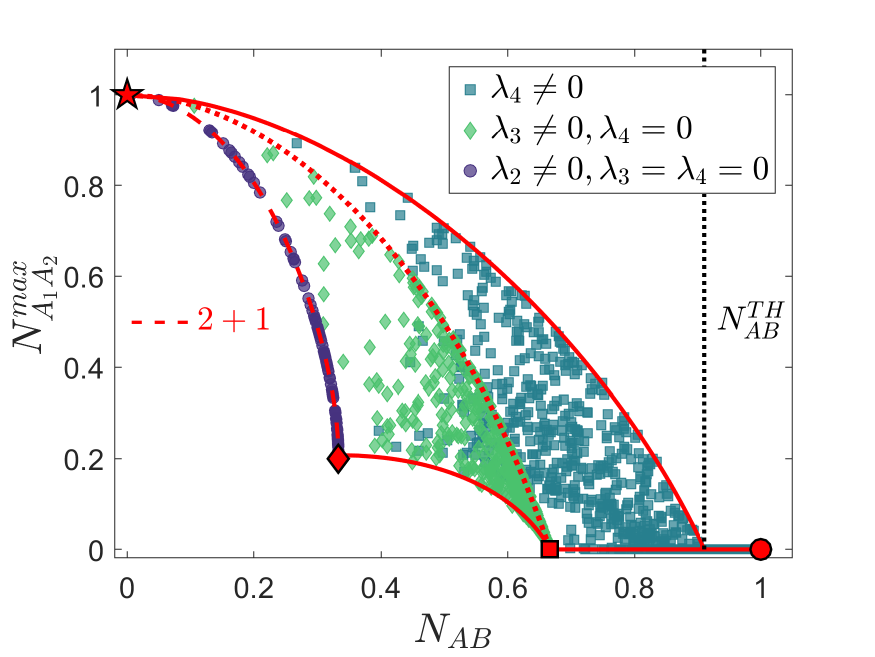}
\caption{$N^{max}_{A_1A_2}$ as a function of $N_{AB}$. The markers represent $3000$ randomly sampled points for a 2+N system. The sampling is done by randomly choosing the four eigenvalues $\lambda_1$, $\lambda_2$, $\lambda_3$, and $\lambda_4$ of the reduced density matrix. Purple circles (below the red dashed line) represent the states for which $\lambda_3=\lambda_4=0$, green diamonds the states where $\lambda_4=0$, and blue squares the states where all eigenvalues are non-zero. Red lines signal the boundaries found through the minimization/maximization of Eq.\eqref{MaxN2Qubits}. Specifically, the red dashed line represents the monotonic expression for a 2+1 system in Eq.\eqref{MaxN2Monotonic} after being re-scaled by a factor of 1/3, while the red dotted line is the result of the maximization subject to the additional constraint $\lambda_4=0$. Red markers signal representative boundary states defined explicitly in the main text, while the vertical dotted line illustrates the threshold value of the negativity.}
\label{Nmax}
\end{figure}

As discussed in the main text, in the $2+N$ case, the relation between $N_{AB}$ and $N^{max}_{A_1A_2}$ is described by a region whose boundaries can be found, for instance, using Lagrange multipliers. We find these boundaries numerically and report them as red lines in Fig.~\ref{Nmax}, \corr{although we note that analytical expressions for these boundaries are provided in Ref.~\cite{Calamet2017}.}

The solid red lines Fig.~\ref{Nmax} represent the results of the minimization/maximization of Eq.~\eqref{MaxN2Qubits} subject to the constraints: (i) $\sum_{i=1}^4 \lambda_i=1$ and (ii) $N_{AB}=N_{AB}^*$ in Eq.~\eqref{2+N}, with $0\leq N_{AB}^*\leq1$ (more details below). If we add the additional constraint $\lambda_3=\lambda_4=0$ to the minimization, we obtain the lower boundary signaled by a red dashed line, which is equivalent to the analytic result obtained for the 2+1 system (Eq.~\eqref{MaxN2Monotonic}) after rescaling by a factor of $1/3$. The endpoints of this boundary are states where A is in a pure state (red star in Fig.~\ref{Nmax}) and states where $\lambda_1=\lambda_2=1/2$ (red diamond), with the GHZ states of the form $\vert \psi \rangle_{\rm GHZ} = \frac{1}{\sqrt{2}}\vert 00 \rangle \otimes \vert 0 0 ...0\rangle +\frac{1}{\sqrt{2}}\vert 11 \rangle \otimes \vert 1 1 ...1\rangle $ \cite{greenberger1990bell,Greenberger1989} being an example of the latter.

A different boundary (red dotted line) is obtained if, instead, the additional constraint $\lambda_4=0$ is considered. This line denotes the maximum value that $N^{max}_{A_1A_2}$ can take for any $\rho_A$ with three non-zero eigenvalues. The vertex where this line connects to the general lower boundaries (red square) represents the family of states for which $\lambda_1=\lambda_2=\lambda_3=1/3$, which turn out to be the only states with $\lambda_4=0$ for which no entanglement can be generated within subsystem A.

To confirm that all possible 2+N states are contained within these boundaries, we sample different states $\rho_{AB}$ by randomly choosing the four eigenvalues $\lambda_1$, $\lambda_2$, $\lambda_3$, and $\lambda_4$ for the reduced density matrix $\rho_A$. Subsequently, we compute the negativity $N_{AB}$ using Eq.~\eqref{2+N} and the maximum negativity within A through Eq.~\eqref{MaxN2Qubits}. The results are presented with different markers in Fig.~\ref{Nmax} and show perfect agreement with the boundaries found through numerical optimization.

Finally, in the region of $\lambda_4\neq 0$, we note that for values of $N_{AB}$ larger than a certain threshold $N_{AB}^{TH}$, no entanglement can be generated within subsystem A. \corr{This entanglement threshold was previously discussed in \cite{Calamet2018}}. To find the expression for the threshold, we solve $\sqrt{(\lambda_1-\lambda_3)^2+(\lambda_2 - \lambda_4)^2}=\lambda_2 + \lambda_4$ (see Eq.~\eqref{MaxN2Qubits}) with the additional constraints $\lambda_1>\lambda_2>\lambda_3>\lambda_4$. The largest value of $\lambda_4$ that satisfies this equation is $\lambda_4^{TH}=1/6$, corresponding to a state of the form $\lambda_1=1/2$, $\lambda_2=\lambda_3=\lambda_4=1/6$. Computing the negativity for such a state, using Eq.~\eqref{2+N}, we find that $N_{AB}^{TH}=\frac{1}{3}+\sqrt{\frac{1}{3}}$, which agrees with the numerical data as displayed in Fig.~\ref{Nmax}. Any state with $\lambda_4 \geq \lambda_4^{TH}$ yields $\sqrt{(\lambda_1-\lambda_3)^2+(\lambda_2 - \lambda_4)^2} \leq \lambda_2 + \lambda_4$, and no entanglement can be generated within A. This region is bounded by the maximally mixed state $\lambda_1=\lambda_2=\lambda_3=\lambda_4=1/4$ signaled by the red circle.

Finally, we comment on the numerical optimization used to find the boundaries on Fig.~\ref{Nmax}. Once Eq.~\eqref{2+N} is defined, it is possible to minimize/maximize $N^{max}_{A_1A_2}$ in Eq.~\eqref{MaxN2Qubits} subject to the different constraints to obtain the red boundaries shown in Fig.~\ref{Nmax}. This can be formulated in the language of Lagrange multipliers. We define the Lagrangian as:

\begin{equation}
\mathcal{L} = N^{max}_{A_1A_2}(\lambda_1, \lambda_2, \lambda_3, \lambda_4) + \sum_j \Lambda_j g_j(\lambda_1, \lambda_2, \lambda_3, \lambda_4)\,.
\end{equation}

Here $N^{max}_{A_1A_2}$ is explicitly given by Eq.~\eqref{MaxN2Qubits},  $\Lambda_j$ are the Lagrange multipliers, and $g_j$ represent the constraints. In our case, there are two algebraic constraints given by:

\begin{eqnarray}
&&g_1 = \lambda_1+\lambda_2+\lambda_3+\lambda_4 -1\,,\nonumber\\
&&g_2 = N_{AB}(\lambda_1,\lambda_2,\lambda_3,\lambda_4) - N_{AB}*\,.
\end{eqnarray}

The first constraint is set by the trace of the density matrix being equal to one. In the second constraint, $N_{AB}^*$ is the specific value of $N_{AB}$ for which we are optimizing, we vary $0\leq N_{AB}^* \leq 1$. Given these algebraic constraints, we can minimize/maximize $ N^{max}_{A_1A_2}$ by solving the following set of equations:

\begin{eqnarray}
\frac{\partial \mathcal{L}}{\partial{\lambda_n}} = 0, \quad n=1,2,3,4\,,\nonumber \\
\frac{\partial \mathcal{L}}{\partial{\Lambda_j}} = 0, \quad j=1,2\,. \label{LagrangeM}
\end{eqnarray}

Additionally, we can impose inequality constraints since we consider $\lambda_1 \geq \lambda_2 \geq \lambda_3 \geq \lambda_4$ in the definition of Eq.~\eqref{MaxN2Qubits}. To incorporate inequality constraints one can generalize the Lagrange multiplier treatment to the so-called
Karush-Kuhn-Tucker (KKT) conditions, see \cite{guler2010foundations}, for example.

In this generalized formalism, the Lagrangian can be expressed as:

\begin{equation}
\mathcal{L} = N^{max}_{A_1A_2} + \sum_j \Lambda_j g_j + \sum_l \tau_l h_l\,,
\end{equation}

where $h_l$ are a set of inequalities defined by $h_l \leq 0$. In the present case we can define three inequalities $h_1= \lambda_2-\lambda_1$,  $h_2 = \lambda_3 - \lambda_2$, and $h_3 = \lambda_4 -\lambda_3$. Additional to the expressions in Eq.~\eqref{LagrangeM}, the KKT formalism incorporates the following conditions:

\begin{eqnarray}
\frac{\partial \mathcal{L}}{\partial{\tau_l}} \leq 0, \quad l=1,2,3\,,\nonumber \\
\tau_l \geq 0, \quad l=1,2,3\,, \nonumber \\
\tau_l h_l = 0, \quad l=1,2,3\,.\label{LagrangeM2}
\end{eqnarray}

The last two conditions are usually referred to as dual feasibility and complementary slackness, respectively. We solve Eqs.~\eqref{LagrangeM} and ~\eqref{LagrangeM2} numerically by imposing the correct range for each $\lambda_n$, for example, $0.25 \leq \lambda_1 \leq 1$. The results are displayed using solid red lines in Fig.~\ref{Nmax}. If we include an additional algebraic constraint $g_3=\lambda_4$ we obtain an extra boundary signaled by the dotted red line. Moreover, incorporating $g_3=\lambda_3$ and $g_4=\lambda_4$ simultaneously leads to the analytical result found for 2+1 systems in Eq.~\eqref{MaxN2Monotonic} after rescaling by a constant factor of 1/3 (dashed red line).

The optimization code used to find these boundaries, and all other codes used to generate the figures in this work, can be found in Ref.~\cite{code_repository}.

\end{document}